\documentstyle[amsmath,aps,draft,overcite]{revtex}
\begin{document}
\title{Multiresolution analysis in statistical mechanics. II. The wavelet transform
as a basis for Monte Carlo simulations on lattices}
\author{Ahmed E.\ Ismail, George Stephanopoulos, and Gregory\ C. Rutledge}
\address{Department of Chemical\ Engineering, Massachusetts Institute of Technology,\\
Cambridge, MA\ 02139}
\date{\today}
\maketitle

\begin{abstract}
In this paper, we extend our analysis of lattice systems using the wavelet
transform to systems for which exact enumeration is impractical. For such
systems, we illustrate a {\em wavelet-accelerated Monte Carlo} (WAMC)
algorithm, which hierarchically coarse-grains a lattice model by computing
the probability distribution for successively larger block spins. We
demonstrate that although the method perturbs the system by changing its
Hamiltonian and by allowing block spins to take on values not permitted for
individual spins, the results obtained agree with the analytical results in
the preceding paper, and ``converge'' to exact results obtained in the
absence of coarse-graining. Additionally, we show that the decorrelation
time for the WAMC is no worse than that of Metropolis Monte Carlo (MMC), and
that scaling laws can be constructed from data performed in several short
simulations to estimate the results that would be obtained from the original
simulation. Although the algorithm is not asymptotically faster than
traditional MMC, because of its hierarchical design, the new algorithm
executes several orders of magnitude faster than a full simulation of the
original problem. Consequently, the new method allows for rapid analysis of
a phase diagram, allowing computational time to be focused on regions near
phase transitions.
\end{abstract}

\section{Introduction}

One of the fundamental challenges in the simulation of very large systems is
balancing the competing aims of accuracy, both numerical and physical, and
computational efficiency. As the number of degrees of freedom, the number of
interactions, and the number of system parameters become large, the
computational cost and storage requirements of any algorithm which models
these systems rapidly become prohibitive. For a system with $N$ degrees of
freedom, the complexity of simulation algorithms is typically $O\left(
N^{2}\right) $, although this can be reduced to $O\left( N\right) $ for
methods incorporating both cell and Verlet lists\cite{Verlet67} and $O\left(
N^{3/2}\right) $ for methods including Ewald sums,\cite{Ewald21} or
increased to $O\left( N^{3}\right) $ or more for quantum methods.\cite
{Frenkel96} Consequently, very large systems are still expensive to
simulate, even with efficient algorithms. One way to reduce the complexity
of such systems is to employ a {\em coarse-graining method}, which
systematically reduces the number of degrees of freedom in the system, and
thereby the overall complexity of the simulation. Coarse-graining techniques
have been developed for both on- and off-lattice calculations. Lattice
simulations have generally relied on coarse-grainings based on
renormalization group theory,\cite{Ma76b,Swendsen79,Curtarolo02} while
off-lattice simulations include coarse-graining techniques as varied as
united-atom models, mapping to lattice models, \cite{Cho97,Doruker97}
dissipative particle dynamics, \cite{Schlijper95} and bead-and-spring
models. \cite{Tschop98,Tschop98b}

At the same time, any benefits obtained from applying a suitable
coarse-graining technique must be weighed against the principal difficulty
of using such a method---inaccurate numerical results, such as a different
critical point. The use of uncontrolled approximations also presents a
problem, since even if a simulation converges to give an answer, we often
cannot use results from those simulations to provide any insight into the
systems about which we are actually interested. In addition, in virtually
all cases the transformation is irreversible:\ we cannot reconstruct our
original system after we have ``evolved'' a coarse-grained system.

The work presented in this paper represents the application of a new
technique to enhance the performance of traditional lattice simulations
using the wavelet transform method. The theoretical foundations of this
technique are outlined in the preceding paper (henceforth denoted as I\cite
{Ismail02}). First developed as an analysis technique which has found wide
acceptance in signal processing, the wavelet transform has not yet been
extensively applied to the field of molecular simulations. Exceptions to
this include the variational work of Best and Sch\"{a}fer,\cite
{Best94,Best94b} applied to statistical field theory, as well as the work of
Huang in studying self-similarity in high-energy physics.\cite{Huang97}

We choose to study the class of Ising lattices with pairwise and external
field interactions, so that the Hamiltonian can be written in the form 
\begin{equation}
{\cal H}=-\sum_{i}h_{i}\sigma _{i}-\sum_{i}\sum_{k}J_{ik}\sigma _{i}\sigma
_{k},  \label{ha}
\end{equation}
where the indices $i$ and $k$ run over all spins in the system$,$ and the
interaction strengths $h_{i}$ and $J_{ik}$ can vary with position in the
lattice. The nearest-neighbor Ising lattice corresponds to $J_{ik}=0$ unless 
$\sigma _{i}$ and $\sigma _{k}$ are neighboring spins on the lattice, in
which case $J_{ik}=J$. Our goal is to construct a phase diagram for such
systems as efficiently as possible. The original problem would be studied
using the Metropolis Monte Carlo (MMC) technique, along with possible
improvements such as the Swendsen-Wang or Wolff spin-cluster algorithms.\cite
{Swendsen87,Wolff89} We introduce a new method for studying this problem,
using the wavelet transform technique to simulate the system hierarchically,
by systematically transforming (\ref{ha})\ to provide expressions for
potentials at coarser scales.

Since the systems under consideration in this paper are lattices, we present
the wavelet transform using the discrete framework, involving filter banks
and matrices. In the discussion below, we present equations for
one-dimensional systems only; however, the simulations are performed using
higher-dimensional wavelets created using the lifting scheme of Sweldens. 
\cite{Sweldens97} The lifting scheme can be applied recursively as needed to
achieve representations of the data at multiple length scales. However, we
do not move between different physical models for the system, but only
change the length scales which the different variables describe. As a
result, the wavelet transform can be described as a {\em multiresolution}
technique;\cite{Mallat89,Mallat89b,Sweldens97,Daubechies98} multiresolution
techniques comprise a subset of the more general class of multiscale
techniques, which can describe any simulation involving multiple modeling
steps, possibly involving the use of multiple underlying physical models. 
\cite{Broughton99,Nieminen02} We refer the reader to paper I for further
details on the implementation of the wavelet transform with respect to
lattice systems.

\section{The Wavelet-Accelerated Monte Carlo (WAMC) algorithm}

The principal difficulty of performing a wavelet transformation on a lattice
system is in working with the discrete set of values that each spin is
permitted to take, such as in a spin-$q$ Ising model. Because the
transformed variables are linear combinations of the original variables, the
constraint that the spins of the individual sites on the original lattice
must be drawn from the set $\left\{ -q,-q+1,\ldots ,q-1,q\right\} $ quickly
becomes a more complicated constraint on the transformed variables $\tilde{s}%
_{i}$ and $\tilde{\delta}_{i}$. As the system size becomes large, the
difficulty of rewriting the spin constraints proves so great that previous
investigations of the use of wavelets in statistical field theory ignored
Ising models altogether. \cite{Best94} Consequently, we would like, if at
all possible, to avoid computations involving original states after we have
carried out the wavelet transformation. At this point, we take note of the
application of wavelets to image compression, where the goal is to reduce
the amount of information needed to reconstruct an image. We would like to
apply this technique to lattice systems, and reduce the number of degrees of
freedom which must be accounted for in our calculations.

We consider our system to be a $d$-dimensional regular lattice ${\cal L}$
with side length $l$, so that the size of the lattice is $N=\left| {\cal L}%
\right| =l^{d}$, and we let a site $\sigma _{i}\ $on the lattice ${\cal L}$
be characterized by a ``spin'' chosen from a finite set ${\cal J}$ of values
and by its physical location on the lattice. For the spin-$\frac{1}{2}$
Ising model, for example, the set ${\cal J}$ is just $\left\{ +\frac{1}{2},-%
\frac{1}{2}\right\} $ (although for computational convenience this is
usually treated as $\left\{ +1,-1\right\} ,$ a convention which we follow
below); similarly, for a lattice gas based on a spin-$1$ Ising model, ${\cal %
J}=\left\{ 0,1,2\right\} $ represents the allowed occupation numbers of each
lattice site. We then assume that the only physical interactions that occur
are either interactions with an external field $h_{i}$ which can vary at
each lattice site, or pairwise interactions with the bilinear form $U\left(
\sigma _{i},\sigma _{j}\right) =J_{ij}\sigma _{i}\sigma _{j}$, where $J_{ij}$
is usually a function only of the spacing between sites $i$ and $j$.
Consequently, the Hamiltonian of the system can be written in the form (\ref
{ha}), 
\begin{equation}
-\beta {\cal H}=\sum_{i}h_{i}\sigma _{i}+\sum_{i}\sum_{j}J_{ij}\sigma
_{i}\sigma _{j}.  \label{h1}
\end{equation}
For the purposes of our simulations, however, we will find it more
convenient to treat the set of spins $\left( \sigma _{1},\ldots ,\sigma
_{N}\right) $ and the external field $\left( h_{1},\ldots ,h_{N}\right) $ as
vectors ${\bf u}$ and ${\bf h}$, and the pairwise interaction strengths $%
J_{ij}$ as a matrix ${\bf J}$. Then the Hamiltonian (\ref{h1}) can be
written in matrix form as 
\begin{equation}
-\beta {\cal H}={\bf h}^{T}{\bf u}+{\bf u}^{T}{\bf Ju}.  \label{h2}
\end{equation}
This formulation of the problem is similar in spirit to that of graph
theory, where the pairwise potential $J_{ij}$ is used to generate an
adjacency list which specifies which edges interact. \cite{Cormen90} Using (%
\ref{h2}) as the basis for a Monte Carlo simulation requires the calculation
of the change of energy $\Delta E_{nm}$ from microstate ${\bf u}_{m}$ to
microstate ${\bf u}_{n}$: 
\begin{equation}
\Delta E_{nm}={\bf h}^{T}\left( {\bf u}_{n}-{\bf u}_{m}\right) +\left( {\bf u%
}_{n}-{\bf u}_{m}\right) ^{T}{\bf J}\left( {\bf u}_{n}-{\bf u}_{m}\right) .
\label{h3}
\end{equation}
If moves are restricted to changes of single spin flips, then only a single
entry of ${\bf u}_{n}-{\bf u}_{m}$ is nonzero, and therefore the calculation
(\ref{h3})\ reduces to a dot product, instead of a matrix multiplication.

As described in I, the action of the wavelet transform is to insert between
each product in (\ref{h2}) or (\ref{h3}) the identity matrix in the form $%
{\bf I}={\bf W}^{T}{\bf W},$ where ${\bf W}$ is the wavelet transform which
maps data from one scale to the next coarser scale, containing half as many
data points. The resulting expressions rewrite the Hamiltonian in terms of
wavelet-transformed averages and differences, with downsampling needed to
reduce the number of variables from $2N$ to $N$. As before, the wavelet
transform can be iterated by applying it to successive sets of averages,
leading after $K$ iterations to Hamiltonians of the form: 
\begin{equation}
-\beta \tilde{H}=\left( {\bf \tilde{h}}^{\left( K\right) }\right) ^{T}{\bf 
\tilde{u}}^{\left( K\right) }+\left( {\bf \tilde{u}}^{\left( K\right)
}\right) ^{T}{\bf \tilde{J}}^{\left( K\right) }{\bf \tilde{u}}^{\left(
K\right) },  \label{h4}
\end{equation}
where in (\ref{h4}) the ${\bf \tilde{u}}^{\left( K\right) }$ represent
``block spins'' whose values are determined by wavelet averaging over some
well-defined region of the original system. The Hamiltonians (\ref{h2})\ and
(\ref{h4})\ have the same formal structure, so that Monte Carlo simulations
of the two systems are essentially identical. The only modifications needed
to simulate a coarse-grained Hamiltonian are the ability to select new
microstates ${\bf \tilde{u}}_{i}^{\left( K\right) }$ which are generated
through wavelet transformations of the original microstates ${\bf u}_{i},$
and the elimination of unwanted degrees of freedom from (\ref{h4}). It
should be noted that in (\ref{h4}), the elements of ${\bf \tilde{u}}^{\left(
K\right) }$ are not restricted to the same values as in the original system,
but are free to take on any value which is consistent with the wavelet
transform applied to the system.

From above, we saw that for even Hamiltonians $\tilde{H}\left( {\bf \tilde{x}%
}\right) $, we should have that $\left\langle \delta \right\rangle _{\tilde{H%
}}=0$ for any wavelet difference $\delta ,$ where $\left\langle \cdot
\right\rangle _{\tilde{H}}$ denotes the ensemble average weighted by the
Hamiltonian $\tilde{H}$. As a ``worst-case scenario'' for our method, we
shall assume not only that $\left\langle \delta \right\rangle _{\tilde{H}%
}=0, $ but also that any terms in the Hamiltonian (\ref{h4}) containing
fluctuation terms can be neglected as well. This assumption allows us to
reduce the size of ${\bf \tilde{J}}^{\left( K\right) }$ from $N\times N$ to $%
2^{-Kd}N\times 2^{-Kd}N$, where $d$ is the lattice dimensionality.
Consequently, instead of performing calculations involving all of the
original variables ${\bf x}$ which describe the state of our system, we
consider functions only of local averages of our original variables.
However, we anticipate that this simplification of the interactions present
in the system will have a significant impact on the thermodynamic behavior
of the resulting system; we illustrate these effects below.

To generate the new microstates ${\bf \tilde{u}}_{i}^{\left( K\right) }$, we
need an estimate for the probability distribution $p\left( \tilde{u}%
_{i}^{\left( K\right) }\right) $ which describes the individual sites in the
coarse-grained lattice. Determining the correct distribution for a given $%
\tilde{u}_{i}^{\left( K\right) }$ would require a detailed simulation of the
original system. An alternative, ignoring the effect of neighboring block
spins, would be to perform an exact enumeration of the spins within a block,
which is possible only for the smallest of block spins. Since we would like
to apply this method to systems of arbitrary size, we want to avoid both of
these options. Therefore, we simulate a sublattice with the same dimensions
as $\tilde{u}_{i}^{\left( K\right) }$, ignoring physical interactions with
the rest of the system by using either free or periodic boundary conditions.
Using the standard Metropolis acceptance criterion, we compute distributions
for the properties of the small lattice, such as the magnetization. Then,
according to the matrix formulation described in Sections II and III of
Paper I,\cite{Ismail02} since the wavelet transform defines a single block
spin $\tilde{u}_{i}^{\left( K\right) }$ as a linear function of the
individual spins at level $K-1$ which it replaces, we can use the linearity
properties of probability distributions to convert the distribution of the
properties directly into a distribution for the block spin $\tilde{u}%
_{i}^{\left( K\right) }$.\cite{Billingsley95} Finally, using the
distribution for the block spin $\tilde{u}_{i}^{\left( K\right) }$ as a
starting point, we perform a Monte Carlo simulation on the system of block
spins defined by the Hamiltonian (\ref{h4}).

Although (\ref{h2})\ and (\ref{h4}) are structurally the same, we cannot
impose a one-to-one correspondence between the states in the configuration
space of (\ref{h3}) and the states in the configuration space of (\ref{h4}).
Consequently, the thermodynamic information obtained from the two will not
necessarily be identical; as we have shown in paper I, there is under fairly
broad conditions a loss of entropy associated with the application of
coarse-graining to a system. We can ensure that the detailed balance
condition for the simulation based on (\ref{h4}) is satisfied for the new
simulation by requiring 
\[
\frac{\alpha \left( {\bf \tilde{u}}_{m}^{\left( K\right) }\rightarrow {\bf 
\tilde{u}}_{n}^{\left( K\right) }\right) }{\alpha \left( {\bf \tilde{u}}%
_{m}^{\left( K\right) }\rightarrow {\bf \tilde{u}}_{n}^{\left( K\right)
}\right) }=\frac{p\left( {\bf \tilde{u}}_{n}^{\left( K\right) }\right) }{%
p\left( {\bf \tilde{u}}_{m}^{\left( K\right) }\right) }e^{-\beta \left( 
{\cal H}\left( {\bf \tilde{u}}_{n}^{\left( K\right) }\right) -{\cal H}\left( 
{\bf \tilde{u}}_{m}^{\left( K\right) }\right) \right) }, 
\]
where $\alpha \left( m\rightarrow n\right) $ is the probability of accepting
a move from microstate $m$ to microstate $n,$ and $p\left( m\right) $ is the
probability of selecting microstate $m$ as determined from simulations on
finer-grained lattices at lower scales.

\section{Theoretical performance of WAMC versus traditional MC}

The wavelet transform is a {\em hierarchical} method which can be applied
iteratively to a system to obtain successively coarser descriptions of a
system. To describe the operation of the wavelet transform on a lattice
model, we need to introduce some notation based on the various length scales
in the problem. In the original problem, the applicable length scales are
the lattice spacing $l$, the correlation length $\xi ,$ and the total
lattice size $L$. Applying the wavelet transform method once increases the
lattice spacing by some factor $a$, so that the ratios of correlation length
to lattice spacing and of system length to lattice spacing each decrease by $%
a$. If we apply the wavelet transform $m$ times in succession, the
corresponding factor becomes $a^{m}$.

We perform the simulation in a series of $K$ stages, where the length scales
at each stage are functions of the length scales at the previous stages. The
initial simulation is performed on a sublattice of the original problem,
with lattice size $L^{\left( 1\right) }<L$, where the superscript denotes
the first stage of the simulation. The lattice spacing of the first stage is
the same as in the original problem, so we define $l^{\left( 1\right) }=l$.
At each subsequent stage of the simulation, the lattice spacing of the $k$th
stage is defined by the recursive relation $l^{\left( k\right) }=L^{\left(
k-1\right) }l^{\left( k-1\right) }$. Since $l^{\left( 1\right) }$ is fixed
to be the lattice spacing of the original lattice, the adjustable parameter
in this relation is the lattice size $L^{\left( k\right) }$ of each stage.
If we assume that the lattice is the same length in all dimensions at every
stage, a single variable in stage $k$ is a block variable representing the $%
\left( L^{\left( k-1\right) }\right) ^{d}$ variables simulated in stage $k-1$%
.

Assuming that the lattice is the same length in all directions both in the
original problem and at every stage in the wavelet-transformed problems,
there are $N_{t}=L^{d}$ lattice variables in the original problem, and $%
N^{\left( k\right) }=\left( L^{\left( k\right) }\right) ^{d}$ lattice
variables in the $k^{th}$ stage of the wavelet-transformed problem. However,
each variable in stage $k$ is a block variable representing the average
behavior of the $\left( L^{\left( k-1\right) }\right) ^{d}$ variables in a
block at stage $k-1$, so the number of {\em total} degrees of freedom
represented at stage $k$ is $N_{t}^{\left( k\right)
}=\prod_{i=1}^{k}N^{\left( i\right) }$, where $N_{t}^{\left( K\right)
}=N_{t}.$ The number of {\em simulated} degrees of freedom is $N_{s}=N_{t}$
for traditional Metropolis Monte Carlo (MMC), but $N_{s}=\sum_{i=1}^{K}N^{%
\left( i\right) }$ for WAMC. Because the running time of Monte Carlo
simulations is usually linear in the number of degrees of freedom being
simulated, the advantage of coarse-graining the system using a wavelet
transform becomes evident. For example, consider an ``original problem'' of
simulating a cubic lattice with $256$ Ising variables on a side. If we
divide the original problem into two stages consisting of cubes of $16$
Ising variables on a side, we reduce the original problem of analyzing $%
256^{3}=\allowbreak 16\,,777,216$ variables to the simpler problem of
analyzing $2\left( 16^{3}\right) =\allowbreak 8192$ variables. Although it
is more difficult to produce a trial configuration in a simulation of the
wavelet-transformed problem than in a simulation of the original Ising
lattice problem, this is more than offset by the reduction in the number of
degrees of freedom being simulated.

\section{Results}

For the purposes of comparison, our ``experimental'' systems are
two-dimensional Ising models of size $32\times 32$, where we have run both
MMC simulations on the full lattice, and WAMC\ simulations at a variety of
resolutions; we shall denote these resolutions using the notation $\left(
x,y\right) $, where $x$ indicates the length of the block size simulated in
the first stage to estimate the probability distribution $p\left( {\bf 
\tilde{u}}^{\left( K\right) }\right) $ to be used in the second stage, and $%
y $ denotes the number of blocks on a side of the lattice in the second
stage of the simulation.

\subsection{Order parameter}

Usually, the property of greatest interest in a simulation of a lattice
system is the order parameter $\eta $. For spin systems, $\eta $ is
generally taken to be the magnitude of the average magnetization, so that $%
\eta =\left\langle m\right\rangle $. [For {\em XY} and Heisenberg models,
and other models where spins are oriented, we generally consider only the
magnitude of the average vector $\eta =\left\langle m\right\rangle
=\left\langle \left| {\bf m}\right| \right\rangle $.] Generally, this is a
very simple property to compute, since the value of the order parameter is
constantly updated during the course of the simulation, and is thus always
available.

For the $32\times 32$ Ising model, the results of a MMC simulation, as well
as $\left( 4,8\right) $- and $\left( 8,4\right) $-WAMC simulations are shown
as Figure \ref{fig3}. The primary difference in the curves for the three
cases is that as the coarse-graining process decreases the number of degrees
of freedom in the final stage of the simulation, the location of the Curie
temperature, indicating onset of spontaneous magnetization, increases and
the steepness of the curve below the Curie temperature decreases. This
result is consistent with our findings for average absolute magnetization $%
\left\langle \left| m\right| \right\rangle $ from analytical models,
discussed in paper I. In the present case, we note further that we achieve
agreement between the different models not only in the low-temperature
region, but also in the high-temperature regime $T\gg T_{c}$. The
differences in the intermediate regime can be attributed largely to the
difference in behavior that results from the use of the wavelet transform to
move from the original Hamiltonian (\ref{h2}) to a coarse-grained
Hamiltonian (\ref{h4}). Additionally, the increased noise in the WAMC
results at intermediate and high temperatures arises because of the
approximations used for the probability distributions $p\left( {\bf u}%
^{\left( K\right) }\right) $ at the second stage of the simulation. The
relative lack of noise in the MMC results stem in part from the fact that
the Metropolis technique leads to non-ergodic sampling of phase space as
temperature increases, as the simulation tends to cycle through a limited
number of states.\cite{Landau00}

\subsection{Internal Energy}

Plotting the internal energy $\left\langle U\right\rangle $ as a function of
the temperature, we obtain curves that follow the same general pattern
outlined in paper I. As illustrated in Figure \ref{fig4}, at low
temperatures, the internal energy, as computed for the $32\times 32$ model
using standard MC as well as $\left( 4,8\right) $- and $\left( 8,4\right) $%
-WAMC\ simulations, is in exact agreement for all methods. This occurs
because only a few microstates of the system, corresponding to states that
have all spins aligned, are actually observed by the system, and the wavelet
transform preserves the energy of these states exactly. All three eventually
reach an average internal energy of zero, but exact agreement is only
expected in the infinite-temperature limit, when the difference in energy
levels between microstates becomes unimportant. For intermediate
temperatures, as before, the disagreement is a result of the change in form
of the Hamiltonian that results from neglecting local correlations. Also, we
note that for WAMC the ``noise'' in the internal energy increases both with
increasing proximity to the ``observed'' critical point of the system as
well as with increasing coarse graining. The additional coarse graining
yields a Hamiltonian with reduced numbers of energy levels, since the energy
of a block spin is defined here to be a function only of its overall
magnetization, and not of its internal magnetization fluctuations; the
reduced number of discrete energy levels yields noisier data.

\subsection{Fluctuation properties}

Fluctuation properties are useful for locating critical points since in the
vicinity of a critical point, the magnitude of fluctuation properties is
known to diverge as $t^{-\alpha }$, where $t\equiv \left| T-T_{c}\right|
/T_{c}$. \cite{Pathria96} Thus, a rapid increase in the value of a
fluctuation property such as the heat capacity at constant external field, $%
C_{H}=\left( \left\langle E^{2}\right\rangle -\left\langle E\right\rangle
^{2}\right) /k_{B}T^{2}$, with respect to temperature can be used to
estimate the critical temperature of a system. However, use of the wavelet
transform leads to a decrease in the magnitude of the heat capacity, since
the coarse-graining leads to smaller variances in the distribution of the
energy $\left\langle U\right\rangle $. Consequently, the maximum value of
the heat capacity $C_{\max }$ decreases as a function of the number of
degrees maintained in the problem.

In Figure \ref{fig5}, the heat capacity is shown as a function of
dimensionless temperature $k_{B}T/J$ for the same systems as for the order
parameter and internal energy measured above. We see that the location of
the maximum of the heat capacity does increase, as expected. Although the
relative maximum of the heat capacity obtained from the $\left( 4,8\right) $%
- and $\left( 8,4\right) $-WAMC\ simulations appears to be identical, they
differ by about 3 percent. Moreover, the actual value of the maximum is not
as important as its existence and its location as a function of $k_{B}T/J$
and of the resolution of the model.

\subsection{Scaling results}

One application of the wavelet-accelerated MC method is to provide an upper
bound for locating phase transitions. Running multiple simulations, at
different levels of resolution, one can determine for each level of
resolution the approximate phase transition temperature $T_{p}\left(
N_{s}\right) $, where $N_{s}$ is the number of degrees of freedom (here,
block spins) in the given system. From these data, one can extrapolate a
scaling relationship of the form 
\begin{equation}
\left( T_{p}-T_{p}^{\ast }\right) \sim N_{s}^{-\gamma },  \label{sr1}
\end{equation}
where $\gamma $ is the corresponding ``scaling'' exponent, and $T_{p}^{\ast
} $ is the phase transition temperature for the untransformed model. The
estimate obtained for the scaling exponent $\gamma $ depends upon the
technique used to calculate the transition temperature for a given
system---for example, estimating divergence of the heat capacity versus the
onset of spontaneous magnetization. Our simulations suggest that $\gamma $
is typically between $0.20$ and $0.25$, with lower values obtained from
divergence of heat capacity than from the onset of spontaneous magnetization.

As explained in I, using a relationship like (\ref{sr1}) to estimate the
phase transition temperature will usually lead to an overestimate of the
phase transition temperature. This is a consequence of the underestimation
of entropy that occurs through the reduction of the size of configuration
space as a result of the wavelet transform. Estimates of $k_{B}T_{p}$/$J$,
as determined by (\ref{sr1}) for the two-dimensional Ising model considered
here typically varied between about $2.7$ and $2.9$, which is an error of
approximately 25 percent from the theoretical value of $2.27$ provided by
the Onsager solution, but only about 20 percent from the results determined
by the traditional MC\ simulations, which gave $k_{B}T_{p}/J\approx 2.35$.
Although these errors are somewhat sizable, it is useful to note that the
total computation time required to obtain the estimate using scaling laws is
at least an order of magnitude smaller than the computation time required to
perform a direct simulation on the original system. Thus, if computational
time is at a premium, an effective approach may be to use the wavelet
transform method to provide an upper bound for $T_{p}$, and then perform a
direct simulation for the parameter space with temperatures below $T_{p}$.

\subsection{Decorrelation time}

Another important measure to study is the time required for decorrelated
samples. It is well known that in the vicinity of the critical point,
traditional Monte Carlo algorithms experience so-called ``critical
slowing-down.'' \cite{Binney93} The Monte Carlo aspect of the WAMC\
algorithm does not vary from traditional Metropolis Monte Carlo, so we
expect that the performance of the two algorithms should be similar, when
measured near their respective critical temperatures.

To compare the two methods, we generated $2^{24}=\allowbreak 16\,777\,216$
new configurations for the $32\times 32$ Ising model at the critical point
using traditional Metropolis Monte Carlo, as well as for the $\left(
4,8\right) $-, $\left( 8,4\right) $-, and $\left( 16,2\right) $-WAMC models.
To determine the correlation time, we used the blocking technique of
Flyvbjerg and Petersen. \cite{Flyvbjerg89} The results are shown as Figures 
\ref{fig1} and \ref{fig2} for the MMC\ and $\left( 16,2\right) $-WAMC
models, respectively. The salient feature in the graph is the onset of a
plateau in the value of the variance of the energy $\sigma _{E}^{2}$;
according to the method of Flyvbjerg and Petersen, this indicates that
configurations separated by a distance of $2^{x}$ steps are statistically
independent, where $x$ is the number of blocking transformations that have
been performed. For the MMC\ model, we find that $x\approx 13$ or $x\approx
14$ provides a decent estimate; for the $\left( 16,2\right) $-WAMC\ model, $%
x=16$ is a good estimate for the index. [For the $\left( 4,8\right) $- and $%
\left( 8,4\right) $-models (not shown), $x=15$ is a reliable estimate.] In
each case, this indicates that between $2^{14}=\allowbreak 16\,384$ and $%
2^{16}=\allowbreak 65\,536$ steps are required between independent
configurations. Thus, we conclude that there is no degradation of
performance near the critical point of a WAMC simulation, relative to
traditional MMC simulations.

\section{Analysis}

\subsection{Measured performance comparison}

In comparing the performance of the standard Monte Carlo algorithm to the
wavelet-accelerated Monte Carlo algorithm, we performed $5\times 10^{5}$
lattice passes on a $32\times 32$-lattice on a 733 MHz Pentium II: for the
standard Monte Carlo algorithm, this meant that, on average, $5\times 10^{5}$
attempts were made to flip each spin. For the WAMC\ algorithm, $5\times
10^{5}$ attempts were made to flip a spin on a given level. The results are
summarized in Table 1. We see that the $\left( 4,8\right) $- and $\left(
8,4\right) $-simulations, which have the smallest total number of lattice
sites ($80$), perform the fastest; however, even the $\left( 16,2\right) $%
-simulation, which has one-fourth as many variables ($260$) as the $32\times
32$ standard Monte Carlo simulation ($1024$), finishes in less than $8$ per
cent of the time required for the latter simulation. As the system size
increases, the computational efficiency achieved by breaking down the system
into multiple stages, all of relatively equal size, becomes even greater:\
for a $128\times 128$-lattice, the performance gain increases from a factor
of approximately 25 to a factor of approximately 50, when we compare the $%
\left( 32,4\right) $- and $\left( 16,8\right) $-WAMC\ simulations to the
standard MC model. However, for the $\left( 8,16\right) $-WAMC model, the
complexity of assigning one of $65$ possible values to each of $256$
variables according to the correct probability distribution becomes
comparable to that of the original problem, so that in fact standard MC\
runs in roughly a factor of 3 faster than the $\left( 8,16\right) $-model.

Similar results are also observed for calculating the phase diagram of a $%
64\times 64$-Ising lattice, which is shown in Figure \ref{fig6} as a plot of
average magnetization as a function of temperature and external field
strength using an $\left( 8,8\right) $-model, for temperatures between $%
T=0.5 $ and $T=5.0$, and for field strengths between $h=-1$ and $h=1$. The
phase diagram reproduces the essential features of the original
two-dimensional ferromagnetic Ising lattice, such as the phase separation at 
$h=0,$ although the exact shape differs from the results obtained via a
standard Metropolis Monte Carlo simulations. However, the plot based on
WAMC\ calculations is created approximately 40 times faster than would a
comparable plot using standard MMC.

\subsection{Comparison with renormalization group methods}

Our observations also indicate that the accuracy of the wavelet-accelerated
Monte Carlo simulations depends on the relative proximity to an ``attractive
fixed point'' of the physical model in parameter space. Borrowed from
renormalization group theory, these attractive fixed points represent the
limiting behavior of the system under various conditions, such as the zero-
and infinite-temperature limits and the limits of zero and infinite external
field. As we approach these limiting cases, the approximations made in
obtaining our wavelet-transformed Hamiltonian become increasingly less
significant.

Combining these observations allows us to design an on-line fine-tuning
algorithm for the coarse graining of our system: the further away from the
critical point of the parameter space, the smaller the number of degrees of
freedom $N^{\left( K\right) }$ necessary to simulate the system must be.
Thus, if we keep track of changes in fluctuation properties such as the heat
capacity or the magnetic susceptibility as we change the system parameters $%
\left( h,T\right) $, we can get an estimate of our relative distance to the
critical point. If we are sufficiently far away from the critical point, we
can choose either to increase the number of stages $K$ that we simulate, or
we can look at more degrees of freedom at lower stages by increasing $%
N^{\left( 1\right) },N^{\left( 2\right) },\ldots ,N^{\left( K-1\right) }$.
As we approach the critical point, we can either decrease the number of
stages $K$ or include more degrees of freedom at higher stages by reducing $%
N^{\left( 1\right) },\ldots ,N^{\left( K-1\right) }$.

\subsection{Sources of error}

In general, the source of our errors can be traced to the assumption that
local fluctuation terms could be reasonably ignored in our coarse-grained
Hamiltonian (\ref{h4}). This naive but otherwise useful assumption yields
correct thermodynamic behavior when the overall physics of the system is
particularly simple: in the low- and high-temperature regimes, for instance,
when the number of observed microstates is small or when the differences
between observed microstates is inconsequential. For more complicated
behaviors, as found at intermediate temperatures and above all in the
vicinity of a critical point, the use of this assumption has a drastic
effect on both the phase space of the system, which in turn affects all
thermodynamic properties of the system, including the internal energy and
entropy of the system, as well as fluctuation properties of the system.

More complicated methods for dealing with fluctuation terms have significant
drawbacks associated with them. Treating the fluctuation terms just like the
block averages maintained in ${\bf \tilde{u}}^{\left( K\right) }$ means that
the wavelet-transformed Hamiltonian is no simpler than the original
Hamiltonian, which affords few advantages in computational time. Likewise,
other approaches, such as parametrizing the probability distribution for the
elements of ${\bf \tilde{u}}^{\left( K\right) }$ using a property like the
energy $E,$ introduce new functional dependencies which cannot be taken into
account using the wavelet transform. Thus, we sacrifice one of the major
advantages of the method---moving from one level to another is achieved
exclusively through use of the wavelet transform. Thus, the most promising
avenue for dealing with fluctuation terms is to develop a probability
distribution for the fluctuation terms via the same approach used to
determine probability distributions for the local averages. Then, using the
probability distribution for the fluctuation terms, we can treat the
discarded terms of the Hamiltonian as a noise term which can be used to
restore some of the entropy that was lost as a result of the coarse-graining
[see paper I for more details]. However, we have presented our results here
to show what can be achieved under ``worst-case'' conditions, without the
use of inverse coarse-graining methods.

Many coarse-graining techniques control errors by fitting the parameters of
a new Hamiltonian to ensure agreement with some known structural information
about the system, such as the radial distribution function.\cite
{Baschnagel00,Akkermans01} For lattice systems, this iterative approach is
reflected in renormalization group theory, and notably the Wilson recursion
method,\cite{Wilson74} which finds the fixed points of the system. As
formulated, WAMC\ creates a coarse-grained Hamiltonian by truncating the
Hamiltonian obtained after application of the wavelet transform. As an
improvement to this, it should be possible to use the wavelet transform to
determine which terms will appear in the Hamiltonian, and then determine the
appropriate parameters to ensure the best fit for some desired property of
the system using an iterative approach.\cite
{Lyubartsev95,Lyubartsev97,Rutledge01}

\subsection{Constructing an adaptive algorithm for MC using the wavelet
transform}

As pointed out above, the wavelet transform method tends to produce
overestimates for the critical point of the system; therefore, if we start
with the high-temperature limit of our algorithm and slowly reduce the
temperature in our simulation, we can observe the movement toward the
critical point by watching various fluctuation parameters, such as the heat
capacity $C_{H}=\left( \left\langle E^{2}\right\rangle -\left\langle
E\right\rangle ^{2}\right) /k_{B}T^{2}$. Near the critical point, we expect
to see a rapid increase in the value of $C_{H}$, consistent with the
expected logarithmic divergence observed in the limit of finite-size
systems. \cite{Frenkel96,Pathria96} If we use the onset of this logarithmic
divergence as an indicator, we can then ``step down'' and use a finer
lattice including more degrees of freedom. This system will naturally better
reflect the physics of our system, particularly in the vicinity of the
critical point. We expect that very near the critical point, we will have to
simulate the system at the original scale, since this will be effectively
the only level which accurately represents the underlying behavior of the
system. However, the region of parameter space where this is necessary is
relatively small compared to the complete parameter space. This is
especially true when we consider that as we proceed below the critical
temperature of the system, the logarithmic divergence of $C_{H}$ will also
vanish. As a result, as we move increasingly far away from the critical
point, we begin to approach the other fixed-point behaviors associated with
the low-temperature limits of the system. Since these are reasonably
well-preserved using the wavelet transformation, we can safely return to
increasingly coarse-grained descriptions of our system as the simulation
proceeds past the critical point.

As an example, we compute the spontaneous magnetization curve for a $%
64\times 64$ Ising lattice in the temperature range $0.5\leq T\leq 10.0$,
with $\Delta T=-0.05$, and choosing as our refinement criterion $\Delta
C_{H}/\Delta T\leq -0.5$, until we reach the finest scale, corresponding to
the original problem. We begin by coarse-graining the system to an $\left(
8,8\right) $-model, where we find that the criterion is triggered only at $%
T=5.1$; we then continue with a $\left( 4,16\right) $-model, down to $T=4.0$%
, at which point the refinement criterion is exceeded. Refining once more,
we proceed with a $\left( 2,32\right) $-model until $T=3.4$, at which point
the threshold is again crossed. Since the next refinement is the original
problem, we proceed at this level of resolution until we have passed the
critical point, so that $\Delta C_{H}/\Delta T$ is positive. As a coarsening
criterion, we select for simplicity the opposite of the refinement
criterion, $\Delta C_{H}/\Delta T\leq 0.5.$ Using this criterion, we find
that we coarsen the model to the $\left( 2,32\right) $-, $\left( 4,16\right) 
$- and $\left( 8,8\right) $-models at temperatures of $T=1.75$, $T=1.65$,
and $T=1.55$, respectively. The rapid coarsening of the model results from
the higher estimates of the critical point in the coarsened models. Since we
are well past the critical point, we expect changes in the heat capacity as
a function of temperature to be relatively small, and thus it is possible to
obtain accurate results from a relatively coarse model. Computationally, the
time required to create this diagram was only 28 per cent that required to
perform a standard Metropolis Monte Carlo simulation with the same number of
steps. Moreover, in the regions that were not simulated using MMC, the
computation time required was just 8 per cent of the time required for MMC.
The resulting plot of magnetization versus temperature, shown as Figure \ref
{fig7}, compares favorably to the analytical solution of Onsager, which is
also shown.\cite{Onsager44}

\section{Conclusions}

The WAMC\ algorithm can dramatically reduce the time required to calculate
the thermodynamic behavior of a lattice system; the trade-off for these
savings in time is in the accuracy of the results obtained, a general
feature of coarse-graining techniques. The error that results is a function
of position in parameter space: the results obtained are generally accurate
in the vicinity of fixed attractors of the system, and decrease as one
approaches critical points of the parameter space. Near critical points,
deviations from results performed on the original lattice system are the
result of coarse-graining the Hamiltonian by eliminating local fluctuation
terms. Consequently, this suggests that a hierarchical simulation which uses
fluctuation properties such as the heat capacity $C_{H}$ to gauge proximity
to critical points in ``real time'' would yield dramatic savings in the
computation time of the behavior of a lattice system over a wide region of
phase space, as regions of space close to fixed attractors would be
simulated at a very coarse scale, with full-detailed simulations reserved
only for regions of parameter space close to critical points.

\section{Acknowledgments}

Funding for this research was provided in part by a Computational Sciences
Graduate Fellowship (AEI) sponsored by the Krell Institute and the
Department of Energy.

% \bibliographystyle{JCP}
% \bibliography{research}

\begin{references}
\bibitem{Verlet67}  {\sc L.~Verlet}, \newblock {\em Phys. Rev.} {\bf 159},
98 (1967).

\bibitem{Ewald21}  {\sc P.~P. Ewald}, \newblock {\em Ann. Phys.} {\bf 64},
253 (1921).

\bibitem{Frenkel96}  {\sc D.~Frenkel} and {\sc B.~Smit}, 
\newblock {\em Understanding Molecular Simulation: From Algorithms to
  Applications}, \newblock Academic Press, San Diego, 1996.

\bibitem{Ma76b}  {\sc S.-K. Ma}, \newblock {\em Phys. Rev. Lett.} {\bf 37},
461 (1976).

\bibitem{Swendsen79}  {\sc R.~H. Swendsen}, \newblock {\em Phys. Rev. Lett.}
{\bf 42}, 859 (1979).

\bibitem{Curtarolo02}  {\sc S.~Curtarolo} and {\sc G.~Ceder}, \newblock {\em
Phys. Rev. Lett.} {\bf 88}, 255504 (2002).

\bibitem{Cho97}  {\sc J.~Cho} and {\sc W.~L. Mattice}, \newblock {\em
Macromolecules} {\bf 30}, 637 (1997).

\bibitem{Doruker97}  {\sc P.~Doruker} and {\sc W.~L. Mattice}, \newblock
{\em Macromolecules} {\bf 30}, 5520 (1997).

\bibitem{Schlijper95}  {\sc A.~G. Schlijper}, {\sc P.~J. Hoogerbrugge}, and 
{\sc C.~W. Manke}, \newblock {\em J. Rheol.} {\bf 39}, 567 (1995).

\bibitem{Tschop98}  {\sc W.~Tsch{\"o}p}, {\sc K.~Kremer}, {\sc J.~Batoulis}, 
{\sc T.~B{\"u}rger},  and {\sc O.~Hahn}, \newblock {\em Acta Polymer.} {\bf
49}, 61 (1998).

\bibitem{Tschop98b}  {\sc W.~Tsch{\"o}p}, {\sc K.~Kremer}, {\sc O.~Hahn}, 
{\sc J.~Batoulis}, and {\sc T.~B{\"u}rger}, \newblock {\em Acta Polymer.}
{\bf 49}, 75 (1998).

\bibitem{Ismail02}  {\sc A.~E. Ismail}, {\sc G.~C. Rutledge}, and {\sc %
G.~Stephanopoulos}, \newblock {\em Submitted to J. Chem. Phys.} (2002).

\bibitem{Best94}  {\sc C.~Best}, {\sc A.~Sch{\"a}fer}, and {\sc W.~Greiner}, %
\newblock {\em Nucl. Phys. B: Proc. Suppl.} {\bf 34}, 780 (1994).

\bibitem{Best94b}  {\sc C.~Best} and {\sc A.~Sch{\"a}fer}, \newblock %
Variational Description of Statistical Field Theories Using {D}aubechies'
Wavelets, \newblock http://xxx.lanl.gov/abs/hep-lat/9402012, 1994.

\bibitem{Huang97}  {\sc D.-W. Huang}, \newblock {\em Phys. Rev. D} {\bf 56},
3961 (1997).

\bibitem{Swendsen87}  {\sc R.~H. Swendsen} and {\sc J.-S. Wang}, \newblock
{\em Phys. Rev. Lett.} {\bf 58}, 86 (1987).

\bibitem{Wolff89}  {\sc U.~Wolff}, \newblock {\em Phys. Rev. Lett.} {\bf 62}%
, 361 (1989).

\bibitem{Sweldens97}  {\sc W.~Sweldens}, \newblock {\em SIAM J. Math. Anal.}
{\bf 29}, 511 (1997).

\bibitem{Mallat89}  {\sc S.~G. Mallat}, \newblock {\em IEEE Trans. Pattern
Analysis and Machine Intelligence} {\bf 11},  674 (1989).

\bibitem{Mallat89b}  {\sc S.~G. Mallat}, \newblock {\em Trans. Amer. Math.
Soc.} {\bf 315}, 69 (1989).

\bibitem{Daubechies98}  {\sc I.~Daubechies} and {\sc W.~Sweldens}, \newblock
{\em J. Fourier Anal. Appl.} {\bf 4}, 247 (1998).

\bibitem{Broughton99}  {\sc J.~Q. Broughton}, {\sc F.~F. Abraham}, {\sc %
N.~Bernstein}, and {\sc \ E.~Kaxiras}, \newblock {\em Phys. Rev. B} {\bf 60}%
, 2391 (1999).

\bibitem{Nieminen02}  {\sc R.~M. Nieminen}, \newblock {\em J. Phys.: Cond.
Matt.} {\bf 14}, 2859 (2002).

\bibitem{Cormen90}  {\sc T.~H. Cormen}, {\sc C.~E. Leiserson}, and {\sc %
R.~L. Rivest}, \newblock {\em Introduction to Algorithms}, \newblock McGraw
Hill-MIT Press, Cambridge, MA, 1990.

\bibitem{Billingsley95}  {\sc P.~Billingsley}, \newblock {\em Probability
and Measure}, \newblock Wiley Interscience, New York, 1995.

\bibitem{Landau00}  {\sc D.~P. Landau} and {\sc K.~Binder}, \newblock {\em A
Guide to {M}onte {C}arlo Simulations in Statistical Physics}, \newblock %
Cambridge University Press, Cambridge, 2000.

\bibitem{Pathria96}  {\sc R.~K. Pathria}, \newblock {\em Statistical
Mechanics}, \newblock Butterworth-Heinemann, Woburn, MA, 1996.

\bibitem{Binney93}  {\sc J.~J. Binney}, {\sc N.~J. Dowrick}, {\sc A.~J.
Fisher}, and {\sc M.~E.~J.  Newman}, 
\newblock {\em The Theory of Critical Phenomena: An Introduction to the
  Renormalization Group}, \newblock Oxford University Press, Oxford, 1993.

\bibitem{Flyvbjerg89}  {\sc H.~Flyvbjerg} and {\sc H.~G. Petersen}, %
\newblock {\em J. Chem. Phys.} {\bf 91}, 461 (1989).

\bibitem{Baschnagel00}  {\sc J.~Baschnagel}, {\sc K.~Binder}, {\sc P.~Doruker%
}, {\sc A.~A. Gusev}, {\sc \ O.~Hahn}, {\sc K.~Kremer}, {\sc W.~L. Mattice}, 
{\sc F.~M{\"u}ller-Plathe}, {\sc M.~Murat}, {\sc W.~Paul}, {\sc S.~Santos}, 
{\sc U.~W. Suter}, and {\sc \ V.~Tries}, \newblock {\em Adv. Poly. Sci.}
{\bf 152}, 41 (2000).

\bibitem{Akkermans01}  {\sc R.~L.~C. Akkermans} and {\sc W.~J. Briels}, %
\newblock {\em J. Chem. Phys.} {\bf 114}, 1020 (2001).

\bibitem{Wilson74}  {\sc K.~J. Wilson} and {\sc J.~Kogut}, \newblock {\em
Phys. Rep.} {\bf 12}, 75 (1974).

\bibitem{Lyubartsev95}  {\sc A.~P. Lyubartsev} and {\sc A.~Laaksonen}, %
\newblock {\em Phys. Rev. E} {\bf 52}, 3730 (1995).

\bibitem{Lyubartsev97}  {\sc A.~P. Lyubartsev} and {\sc A.~Laaksonen}, %
\newblock {\em Phys. Rev. E} {\bf 55}, 5689 (1997).

\bibitem{Rutledge01}  {\sc G.~C. Rutledge}, \newblock {\em Phys. Rev. E}
{\bf 63}, 021111 (2001).

\bibitem{Onsager44}  {\sc L.~Onsager}, \newblock {\em Phys. Rev.} {\bf 65},
117 (1944).
\end{references}

\begin{figure}[tbp]
\caption{Absolute average magnetization as a function of the dimensionless
temperature $k_{B}T/J$ for the $32\times 32$ Ising model computed using
standard MC\ (left curve), a $\left( 4,8\right) $-WAMC\ simulation (center),
and a $\left( 8,4\right) $-WAMC\ simulation (right).}
\label{fig3}
\end{figure}

\begin{figure}[tbp]
\caption{Internal energy as a function of the dimensionless temperature $%
k_{B}T/J$ for the $32\times 32$ Ising model computed using standard MC\
(left curve), a $\left( 4,8\right) $-WAMC\ simulation (center), and a $%
\left( 8,4\right) $-WAMC\ simulation (right).}
\label{fig4}
\end{figure}

\begin{figure}[tbp]
\caption{Heat capacity as a function of the dimensionless temperature $%
k_{B}T/J$ for the $32\times 32$ Ising model computed using standard MC\
(left curve), a $\left( 4,8\right) $-WAMC\ simulation (center), and a $%
\left( 8,4\right) $-WAMC\ simulation (right).}
\label{fig5}
\end{figure}

\begin{figure}[tbp]
\caption{Graph showing variance in the estimate of energy as calculated
using the method of Flyvbjerg and Petersen \protect\cite{Flyvbjerg89} for a $%
32\times 32$ Ising model measured at its critical temperature.}
\label{fig1}
\end{figure}

\begin{figure}[tbp]
\caption{Graph showing variance in the estimate of energy as calculated
using the method of Flyvbjerg and Petersen \protect\cite{Flyvbjerg89} for a $%
32\times 32$ Ising model in a $\left( 16,2\right) $-WAMC measured at the
critical temperature determined from the simulation.}
\label{fig2}
\end{figure}

\begin{figure}[tbp]
\caption{Phase diagram plotting average magnetization versus temperature and
external field strength for a $64\times 64$ ferromagnetic Ising lattice,
computed using an $\left( 8,8\right) $-model via WAMC. The general features
correspond to those that would be produced with standard MMC, but require
less than 3 per cent of the computational time.}
\label{fig6}
\end{figure}

\begin{figure}[tbp]
\caption{Phase diagram plotting average magnetization versus temperature for
a $64\times 64$ ferromagnetic Ising lattice, created using an adaptive WAMC\
algorithm, with refinement and coarsening criterion established using the
change in heat capacity with respect to temperature $\Delta C_{H}/\Delta T$.
The squares represent the simulation results, while the line reproduces
Onsager's analytical result for the two-dimensional Ising model with zero
external field. The simulation used in a given temperature region is shown
on the plot.}
\label{fig7}
\end{figure}

\newpage

Table 1. Performance comparison for Metropolis Monte Carlo (MMC)\ versus
Wavelet-Accelerated Monte Carlo (WAMC) algorithms

\begin{tabular}{|l|l|}
\hline
Simulation & Time for $5\times 10^{5}$ passes (s) \\ \hline\hline
$32\times 32$ MMC & 1824.22 \\ \hline
$\left( 4,8\right) $-WAMC & 64.5781 \\ \hline
$\left( 8,4\right) $-WAMC & 55.1094 \\ \hline
$\left( 16,2\right) $-WAMC & 144.516 \\ \hline
\end{tabular}

\end{document}